# The I-Pu-Xe age of the Moon-Earth system revisited

G. Avice[1] & B. Marty[1]

[1]CRPG-CNRS, Université de Lorraine, 15 rue Notre-Dame des Pauvres, BP 20, F-54501 Vandoeuvre-lès-Nancy Cedex, France

Author for correspondence: G. Avice, gavice@crpg.cnrs-nancy.fr

**Summary**

From iodine-plutonium-xenon isotope systematics, we re-evaluate time constraints on the early evolution of the Earth-atmosphere system and, by inference, on the Moon-forming event. Two extinct radioactivites ($^{129}$I, $T_{1/2}$ = 15.6 Ma, and $^{244}$Pu, $T_{1/2}$ = 80 Ma) have produced radiogenic $^{129}$Xe and fissiogenic $^{131-136}$Xe, respectively, within the Earth, which related isotope fingerprints are seen in the compositions of mantle and atmospheric Xe. Recent studies of Archean rocks suggest that xenon atoms have been lost from the Earth's atmosphere and isotopically fractionated during long periods of geological time, until at least the end of the Archean eon. Here we build a model that takes into account these results. Correction for Xe loss permits to compute new closure ages for the Earth's atmosphere that are in agreement with those computed for mantle Xe. The minimum Xe formation interval for the Earth-atmosphere is $40_{-10}^{+20}$ Ma after start of solar system formation, which may also date the Moon-forming impact.

# 1. Introduction

The age of the solar system is well established at 4.568 Ga [1–3]. Extant and extinct radioactivity systems indicate that not only primitive bodies, but also differentiated planetesimals and planetary embryos including Mars formed within a few Ma after start of condensation in the solar system (inferred from the age of calcium-aluminium rich inclusions - CAIs - in primitive meteorites). In contrast, the formation age of the Earth-Moon system is uncertain and is presently debated within a time interval of 30-200 Ma after CAI (this volume). Deciphering details of the early chronology of the Earth requires the development of adequate extinct radioactivity chronometers. Because the Earth's interior has been well mixed by mantle convection over 4.5 Ga, most of the early reservoirs have been re-homogenized even if some remnants of past heterogeneities might be still present [4–7]. However, information on ancient reservoirs is still kept at the Earth's surface, in old terranes, and, in the case of noble gases, in the terrestrial atmosphere.

Xenon, the heaviest noble gas, has a large number (9) of isotopes, and extant and extinct radioactivity products have contributed several of them. Iodine-129 decays with a half-life of 15.7 Ma into $^{129}Xe$ [8], resulting in $^{129}Xe$ excesses in primitives meteorites relative to the potential primordial Xe isotopic compositions [9]. Atmospheric Xe presents a monoisotopic excess of $^{129}Xe$ (compared to adjacent $^{128}Xe$ and $^{130}Xe$ isotopes) of about 7% (e.g. [10]) attributed to the decay of extinct $^{129}I$. Some natural gases and mantle-derived rocks [11–13] have $^{129}Xe/^{130}Xe$ ratios (where $^{130}Xe$ is a stable isotope of xenon which is used for normalization) higher than the atmospheric value. Altogether, these observations demonstrate that the Earth formed and differentiated while $^{129}I$ was still present, thus within a few tens of Ma. Most (>80 %) of terrestrial Xe is now in the atmosphere (e.g. [11] but see [14] for an alternate view), so that atmospheric Xe is to a first order representative of total terrestrial Xe. Consequently, a $^{129}I$-$^{129}Xe$ age of the Earth can be constrained from estimates of the initial abundance of iodine, inferred from the present-day abundance of the stable isotope $^{127}I$ [15]. Although the latter is not well known (probably not better than a factor of 2, see below), the exponential nature of radioactive decay makes the result not so sensitive to this uncertainty. Thus the I-Xe age of the Earth's atmosphere, which is in fact the time interval $\Delta t_{129}$ of reservoir closure, can be expressed as:

$$\Delta t_{129} = \frac{1}{\lambda_{129}} \ln \left(\frac{^{129}I_{INI}}{^{129}Xe(I)}\right) \qquad (1)$$

where $\lambda_{129}$ is the decay constant of $^{129}$I (4.41x10$^{-2}$ Ma$^{-1}$), $^{129}$I$_{INI}$ is the initial $^{129}$I abundance in the reservoir. The latter is computed with the ($^{129}$I/$^{127}$I)$_{INI}$ initial ratio from meteorite data (1.1x10$^{-4}$) [16] and estimates of terrestrial $^{127}$I abundance (>3 ppb, up to 13 ppb, see next section). $^{129}$Xe(I) represents the $^{129}$Xe excess resulting from the decay of $^{129}$I in the atmosphere (2.8x10$^{11}$ mol. of $^{129}$Xe(I) [10]). Within these assumptions, the Earth would have become closed for Xe isotope loss at 100-120 Ma after CAI, this range depending mostly on the initial abundance of iodine (for further discussion of these parameters, see reviews by [10,15,17–19]). This is the classical "age" of the atmosphere found in textbooks.

The other short-lived nuclide of interest here is $^{244}$Pu (half-life of 80 Ma [20]) which, in addition to α-decay, presents a weak (0.125 %) branch for spontaneous fission and produces $^{131,132,134,136}$Xe isotopes (represented hereafter by $^{136}$Xe(Pu)). These heavy Xe isotopes are also produced in Earth by the spontaneous fission of extant $^{238}$U. However the contribution of $^{238}$U fission to Xe isotopes was minor compared to that of $^{244}$Pu during the periods of time characterizing the Earth's formation and its early evolution. Contrary to iodine, plutonium has no stable isotope, so that the initial abundance of $^{244}$Pu is inferred from comparison with U in meteorites and Earth [21], since both are refractory and lithophile elements. Fissiogenic Xe from $^{244}$Pu has been found in the Earth's interior [4,12,22]. The detection of fissiogenic Xe in the atmosphere is, however, not straightforward, as the original composition of atmospheric xenon is not directly measurable. In fact, atmospheric xenon is isotopically fractionated by 3-4 % per atomic mass unit relative to potential primordial candidates [23]. Furthermore, even after correction for such mass-dependent isotope fractionation, neither chondritic nor solar Xe can be directly related to atmospheric Xe, because both chondritic and solar Xe are rich in the heavy Xe isotopes ($^{134}$Xe and $^{136}$Xe) compared to "unfractionated" atmospheric Xe [24]. Takaoka [25], and Pepin and Phinney [26] extrapolated, from meteorite data, a primordial Xe component (labelled Xe-U by [26]), from which atmospheric Xe could be derived by mass-dependent isotopic fractionation. Xe-U has an isotope composition close to solar Xe for its light masses but is depleted in $^{134}$Xe and $^{136}$Xe. This U-Xe component has still not been found in meteorites, possibly because of the presence of superimposed components of nucleosynthetic origin. Indeed xenon trapped in different meteoritic phases presents variations in its s-, p- and r- process isotopes (e.g., the P3, P6 and HL components trapped in nanodiamonds [27,28]). Thus the heavy Xe isotope difference between potential Xe ancestors and solar Xe could be the result of different mixes of nucleosynthetic Xe isotopes in primitive reservoirs, with the possibility that solar Xe was more contributed by s-process isotopes compared to other Xe primordial progenitors [29]. Whatever the composition of the progenitor of atmospheric Xe, this reservoir appears poor in $^{244}$Pu-derived Xe isotopes.

Estimates of $^{136}$Xe excess in the atmosphere due to contribution of extinct $^{244}$Pu vary between 4.6 % and 2.8 %, according to Pepin and Phinney [26] and Igarashi [30], respectively. $^{136}$Xe(Pu) gives another possibility to estimate closure ages for the Earth-atmosphere. Both the I-Xe and Pu-Xe systems can be combined yielding an I-Pu-Xe time interval $\Delta t_{129-244}$ of:

$$\Delta t_{129-244} = \frac{1}{\lambda_{244}-\lambda_{129}} \ln\left[\frac{^{129}Xe(I)/^{136}Xe(Pu)}{\left(^{129}I/^{244}Pu\right)_{INI}} {}^{136}Y_{244}\right] \qquad (2)$$

where $\lambda_{244}$ is the decay constant of $^{244}$Pu (8.45x10$^{-3}$ Ma$^{-1}$), and $^{136}Y_{244}$ is the production yield of $^{136}$Xe from $^{244}$Pu fission (7x10$^{-5}$, [31]). The $^{129}$Xe(I)/$^{136}$Xe(Pu) ratio of the atmosphere has been estimated to be 4.6 [32], which yields an atmospheric $\Delta t_{129-244}$ closure time of about 100 Ma after CAI, consistent with the I-Xe age. The fact that these two closure ages are comparable is not merely a coincidence since $\Delta t_{129-244}$ depends in large part on the residual amount of $^{129}$Xe(I) in Earth's atmosphere (of the order of 1 %) rather than on that of $^{136}$Xe(Pu) in this time interval, given the much shorter half-life of $^{129}$I compared to that of $^{244}$Pu. A more interesting constraint arises form a direct comparison of the amount of $^{136}$Xe(Pu) left in Earth (atmosphere) to that potentially produced by initial $^{244}$Pu. Although the latter has also a significant uncertainty, it appears that a large fraction of $^{136}$Xe(Pu) (≥70 %) is missing in the present-day Earth's atmosphere [17,32–34]. Given the half-life of $^{244}$Pu of 80 Ma, this discrepancy suggests that Xe was lost from the atmosphere after the giant impact phase of the Earth's accretion [17,33].

Atmospheric xenon is not only isotopically fractionated (enriched in heavy isotopes by 3-4 % per atomic mass unit) compared to potential primordial Xe, but is also elementally depleted by one order of magnitude relative to other noble gases (e.g., Kr) compared to the abundance pattern of meteoritic noble gases. These dual characteristics, known for long as the "xenon paradox" (elementally depleted in heavy element, isotopically enriched in heavy isotopes) have not yet found a satisfactory explanation. The Xe depletion and the lack of Pu-produced Xe isotopes in the atmosphere suggest that, after a last giant impact event, there might have been more xenon in the atmosphere, which would have been lost through a process that fractionated Xe isotopes (if both features were related). Therefore the I-Pu-Xe ages should be corrected for Xe loss and the scope of the correction would depend on the timing of loss relative to I and Pu decays. The apparent deficiency of $^{136}$Xe(Pu) in the atmosphere (see previous paragraph) may indeed be a consequence of prolonged selective loss of atmospheric Xe.

Recent studies of noble gases in Archean (3.5-3.0 Ga-old) rocks may provide a solution to the xenon paradox. Isotopically fractionated Xe has been found in Archean barite [35,36] and hydrothermal quartz [37,38]. The Xe isotopic spectrum is intermediate between the primordial and the modern atmospheric Xe isotope patterns, and the isotopic fractionation (relative to the modern composition) tends to decrease with decreasing age (Fig. 1). Together with Xe data from ancient basement fluids of presumed Proterozoic age [39], the evolution of Xe isotopic fractionation with time is consistent with a Rayleigh distillation in which Xe has been lost from the atmosphere with an instantaneous fractionation factor of about 1.1 % per atomic mass unit [38]. The magnitude of the latter is in agreement with experimental studies of Xe isotope fractionation upon ionization [40]. The exponential decrease of Xe isotope fractionation with time is qualitatively consistent with that of the far UV light (FUV) flux from the evolving Sun with time ([41]; Fig. 1) suggesting that Xe was selectively ionized and lost from the atmosphere to space through time [42,43] at a rate that followed the declining FUV flux.

In this study, we investigate the possibility to reproduce the current features of atmospheric Xe (elemental and isotopic compositions), taking into account this long-term escape. We develop a 3-box model (solid Earth, atmosphere, space) that allows us to correct the abundances of radiogenic/fissiogenic Xe isotopes for Xe loss. Previous computed ages are therefore not valid anymore and their values and meaning have to be revisited. Doing so, we follow Podosek & Ozima [17] who predicted that "If allowance is made for the possibility that most of the Xe, including radiogenic Xe, that should be in the atmosphere, has somehow be removed or hidden, the I-Xe and Pu-Xe formation interval could be reduced to perhaps 60 Ma". We now have observational evidence for such Xe loss through geological time. Since the atmosphere is probably very sensitive to impact-driven erosion [44,45], corrected closure ages may be related to the end of the giant impact epoch that led to the formation of the Moon [46].

## 2. Building of the model

The model consists of three reservoirs: the silicate Earth, the atmosphere and the outer space (Fig. 2). We aim to estimate the closure time of the atmosphere Δt, defined here at the time after CAI when the atmosphere became closed to volatile loss (except for Xe preferentially lost during the Hadean and the Archean eons). Between time 0 (CAI) and Δt, volatile elements are contributed to the proto-Earth by accreting bodies, and partially lost through collision and atmospheric erosion. Between Δt and Present, only Xe is lost to space, the other volatile elements being conservative in the atmosphere. We correct for this secondary Xe loss using the depletion of xenon relative to other noble gases in the atmosphere.

Xenon is degassed without isotopic fractionation from the Earth's interior to the atmosphere through magmatism. Between Δt and the end of the Archean eon, xenon escapes from the atmosphere to the outer space, and is isotopically fractionated during this escape. Three radioactive systems are involved: $^{129}$Xe produced by the β-decay of $^{129}$I ($T_{1/2}$=15.6 Ma), and $^{131,132,134,136}$Xe from the fission of $^{244}$Pu ($T_{1/2}$=80 Ma) and $^{238}$U ($T_{1/2}$=4.47 Ga). As $^{136}$Xe, compared to other xenon isotopes ($^{131,132,134}$Xe), is a major product from the fission of $^{244}$Pu, it will be considered as a proxy for the entire fission component in the following discussion.

The following mass balance exemplifies the evolution of the atmospheric $^{129}$Xe$_{ATM}$ (mol.) with time:

$$\frac{d\,^{129}Xe_{ATM}(t)}{dt} = \varphi(t)\,^{129}Xe_{MANT}(t) - \beta(t)(1 + \alpha_{esc})\,^{129}Xe_{ATM}(t) \qquad (3)$$

where $^{129}$Xe$_{ATM}$(t) is the abundance of $^{129}$Xe in the atmosphere at time t. $^{129}$Xe$_{MANT}$(t) represents the abundance of $^{129}$Xe atoms in the mantle at time t. φ(t) and β(t) are the degassing and escape parameters, respectively. $\alpha_{esc}$ is the isotopic fractionation factor described below (Eqn. 6). The evolution of the mantle $^{129}$Xe$_{MANT}$ (mol.) with time is expressed by:

$$\frac{d\,^{129}Xe_{MANT}(t)}{dt} = -\varphi(t)\,^{129}Xe_{MANT}(t) + \lambda_{129}\,^{129}I(t) \qquad (4)$$

where φ(t) is the degassing parameter at time t, $\lambda_{129}$ is the decay constant of $^{129}$I into $^{129}$Xe, $^{129}$I(t) the abundance of iodine-129 in the mantle at the time t. Similarly, equations for

$^{124-136}$Xe are defined taking into account the decays of the different radioactive nuclides ($^{244}$Pu, $^{238}$U). For example, for $^{136}$Xe$_{MANT}$, the equation is:

$$\frac{d^{136}Xe_{MANT}(t)}{dt} = -\varphi(t){}^{136}Xe_{MANT}(t) + \lambda_{244}B_{244}{}^{136}Y_{244}{}^{244}Pu(t) \quad (5)$$

$$+\lambda_{238}B_{238}{}^{136}Y_{238}{}^{238}U(t)$$

where $^{136}$Xe$_{MANT}$(t) is the abundance of $^{136}$Xe in the mantle at time t, $^{244}$Pu(t) and $^{238}$U(t) are the abundances of parent nuclides in the mantle at time t, B$_{244}$ (1.25x10$^{-3}$) and B$_{238}$ (5.45x10$^{-7}$) are the branching ratios for $^{244}$Pu and $^{238}$U respectively, $^{136}$Y$_{244}$ (6.3 %) and $^{136}$Y$_{238}$ (5.6 %) are the yields of fission [31].

Equations are resolved with 1 Ma-step using an original code written with the Mathematica® programming language. Results of the model comprise, for each temporal step, the amount of each stable isotope in each reservoir plus the amount of each radiogenic/fissiogenic isotope (*e.g.* $^{136}$Xe in the atmosphere coming from the fission of $^{244}$Pu).

*(a) Degassing from the Earth's interior*

The rate of Xe degassing (φ(t) in equations (3-5)) from the Earth's interior through time can be anticipated from thermal and geochemical considerations. Having a constant degassing parameter φ(t) through Earth's history would result in a too severe Xe loss from the mantle. We have tested different functions for φ(t) and model results that best fit the data are those obtained using three different values (φ$_1$ > φ$_2$ > φ$_3$) for the respective intervals of time [Δt, 100 Ma], [100 Ma, 1000 Ma], [1000 Ma, 4500 Ma] (Table 1). An exponential decrease of φ(t) could also fit the data but not as well as this stepped function.

The choices of these steps and of the related time intervals have some physical ground. Due to a higher thermal regime of the solid Earth the degassing rate during the Hadean eon was probably an order of magnitude higher than the modern one [18,47–49]. During the Archean eon, the degassing rate was also probably higher than at present, as indicated for example by the ubiquitous presence of komatiitic lavas, presumably originating from a mantle hotter than today (although some authors argued that a hotter mantle does not necessarily imply an higher convection rate [14,50]). Isotopic fractionation of xenon during magma generation and degassing could only be kinetic, if any, and is neglected here. The model is built in a way that the mantle is degassing Xe into the atmosphere from the time of Earth's accretion, which is mathematically equivalent, during the short time interval of a few tens of Ma, to add Xe from impacting bodies directly to the atmosphere.

*(b) Loss of xenon to the outer space*

As introduced above, Archean rocks of surficial origin contain mass-fractionated Xe isotopically intermediate between Chondritic/Solar and modern Atmospheric (Fig. 2). Data are scarce because they are extremely difficult to obtain (due to the need to date confidently the host phases). Available data are consistent with a time evolution of the Xe isotopic fractionation, presumably in the ancient atmosphere. This evolution can be fitted with a Rayleigh distillation model in which Xe isotopes are escaping from the atmosphere through time with mass-dependent instantaneous isotopic fractionation. The exponential evolution of the isotopic composition of atmospheric xenon with time predicted by Rayleigh distillation is consistent with the decline of FUV light flux from the ancient Sun (Fig. 1), suggesting that interactions between atmospheric Xe and FUV light from the Sun played a role in Xe escape. The Rayleigh distillation equation can be written as:

$$\frac{(\frac{^{i+1}Xe}{^{i}Xe})_{t=4.56\,Ga,today}}{(\frac{^{i+1}Xe}{^{i}Xe})_{t=0}} = f^{\alpha_{esc}-1} \sim 1.035 \qquad (6)$$

where the factor 1.035 relates to the isotopic difference between Solar/Chondritic and modern air and $f$ is the depletion factor of Xe in modern air corresponding to a factor of ≈20 relative to carbonaceous chondrites [23]. The instantaneous fractionation factor $\alpha_{esc}$ is then ≈1.011 % per atomic mass unit. This isotope fractionation is large for an inert gas with such a high mass. Thus either Xe isotopic fractionation resulted from a specific process during atmospheric escape, which is not yet documented, or it involved ionization of xenon, which, from laboratory experiments, has been shown to yield isotopic fractionation of the order of 0.8-1.6 % per atomic mass unit [40,51–55]. Among most volatile species that were potentially present in the ancient atmosphere (e.g., noble gases, CO, $CO_2$, $N_2$, $CH_4$...), xenon has the lowest ionization potential. Hébrard and Marty [42] have investigated the possible behaviour of atmospheric Xe, taking into account the inferred distribution of FUV light wavelengths of the ancient Sun. They proposed that Xe was ionized at an atmospheric height comparable to that of organic haze formation in the Archean atmosphere, so that ionized heavy Xe isotopes were preferentially retained in the lower atmosphere while ionized light Xe isotopes could escape from the upper atmosphere. This possibility is certainly not unique and other processes may be explored but, whatever the origin of this isotopically fractionating Xe loss, it remains firm that Xe isotopically intermediate between primordial Xe and modern atmospheric Xe has been found trapped in Archean rocks.

In our model, we use an instantaneous fractionation factor $\alpha_{esc}$ allowed to vary within 1.0-1.5 % per atomic mass unit. The rate of escape of xenon atoms is scaled using FUV decay curves corresponding to the wavelength at which xenon atoms are ionized (102.3 nm) [41]. The intensity of escape ($\beta(t)$ in Ma$^{-1}$) with time t is given by:

$$\beta(t) = \frac{1}{d}(b \times t)^c \quad (7)$$

b and c are constant parameters from [41] and d is an adjusted constant. Parameters values are shown in Table 1 and the decay of atmospheric Xe with time is shown in Fig. 2.

## 3. Key parameters

The key parameters of the model are either taken directly from the literature (e.g. ($^{129}$I/$^{127}$I)$_{INI}$) when the value is widely accepted or are adjusted testing a range of values when the value is badly known (e.g. I$_{INI}$). Table 1 contains canonical values for parameters of the model and adjusted values.

*(a) Iodine content of the Earth*

Because of its volatility, the precise iodine content of the Earth, and thus the $^{129}$I$_{INI}$, is not well known. A large range between 3 to 13 ppb is proposed in the literature [56–58]. The lower value of 3 ppb is computed with iodine data from the depleted mantle [58] which, by definition, is poor in incompatible elements like iodine. Thus it represents a lower limit of the iodine budget of the Earth. Here we use a value of 6.4 ppb in our reference solution, which gives model results consistent with observed Xe data, and permits a good match between the two chronometers (I-Xe and I-Pu-Xe). The range of I$_{INI}$ values is used as an uncertainty and is propagated in the age calculation. An abundance of 6.4 ppb for $^{127}$I together with the initial $^{129}$I/$^{127}$I ratio of 1.1x10$^{-4}$ obtained from meteorite data [16] yields an initial amount of terrestrial $^{129}$I ($^{129}$I$_{INI}$) of 3.8x10$^{12}$ mol.

*(b) U-Pu content of the Earth*

The range of present-day abundances for $^{238}$U is between 16 and 20 ppb [18] corresponding to initial values, 4.57 Ga ago, between 33 and 41 ppb. Here we take a value of 38 ppb for the initial $^{238}$U content as in [18]. The average value of $6.8 \times 10^{-3}$ for the $(^{244}Pu/^{238}U)_{INI}$ is derived from the analysis of meteorites [59,60] and is consistent with data obtained from ancient terrestrial zircons [18,61–63].

*(c) Initial amount and isotopic composition of xenon*

Xenon is under-abundant in the terrestrial atmosphere relative to other noble gases by a factor of 23±5 relative to Kr [23]. Thus the present-day atmospheric inventory ($6.15 \times 10^{11}$ mol. of $^{130}$Xe [31]) can be corrected for selective escape of xenon atoms by multiplying the current abundance by this factor, which leads to an initial $^{130}$Xe amount of $1.41(\pm 0.36) \times 10^{13}$ mol. It is beyond the scope of this study to evaluate the starting isotopic composition of primordial xenon and we refer to discussions in [17,26,32]. The U-Xe [23] is, so far, the only initial isotopic composition permitting to reproduce the current isotopic composition of atmospheric Xe including fissiogenic and radiogenic contributions [64] and its composition is adopted as a starting isotopic composition for the model.

*(d) Optimisation of the parameters*

The model is constrained using the initial Xe amount of Earth and the U-Xe composition on one hand, and the present-day amount and composition of atmospheric Xe and, for some of the isotopic ratios, of mantle Xe on another hand. The silicate Earth contents of parent elements (I and Pu) are additional parameters that can be adjusted within a plausible range of values. The model must yield ancient Xe isotopic compositions that are consistent with data from Archean rocks.

Key parameters from the literature are $^{130}Xe_{INI}$, $(^{129}I/^{127}I)_{INI}$, $^{238}U_{INI}$, $(^{244}Pu/^{238}U)_{INI}$ and the starting isotopic composition (U-Xe). Other parameters are adjusted following multiple runs. $^{130}$Xe, the stable isotope of reference, is used to scale the degassing and escape rates over time. The different degassing rates $\varphi_1$, $\varphi_2$ and $\varphi_3$ (Table 1) are fitted in order to respect the $^{130}$Xe depletion over time and to take into account variations in the degassing rate of the mantle. The $\varphi_1$, $\varphi_2$ and $\varphi_3$ values that fit best the data are 890, 15 and 1 times the modern rate between $\Delta t$ and 100Ma, 100 Ma and 1 Ga, and 1 Ga and Present, respectively. The variation of the atmospheric Xe escaping rate (Eqn. 7) is scaled to the variation of the FUV light flux over time. Once the degassing and escaping rates of $^{130}$Xe are scaled, it is possible to optimize other key parameters such as the instantaneous fractionation factor $\alpha_{esc}$ and the initial iodine abundance $I_{INI}$. The aim of this final step of optimization is to reproduce the modern and the Archean isotopic compositions of atmospheric Xe within better than 1% (Fig. 2).

*(e) Outcomes of the model*

In addition to optimisation of the degassing and escaping parameters, the model allows one to determine the atmospheric closure time $\Delta t$ that fits best the observations. $\Delta t$ corresponds to the time after CAI when the atmosphere became closed (except for Xe which escape to space continued over eons).

## 4. Discussion

The long-term escape affects the global budget of xenon atoms in the Earth's atmosphere and therefore the amount of radioactive products, which have to be corrected as shown in Table 2.

*(a) Corrected I-Xe age of the Earth-atmosphere*

The I-Xe closure age of the atmosphere, corrected for long-term escape of atmospheric Xe, becomes 41 Ma after CAI, instead of 110 Ma, with our reference I content of 6.4 ppb. For a possible range of 3-13 ppb for terrestrial iodine [56–58], the range of closure age becomes 21-62 Ma (see ranges of solutions in Fig. 3). It must be noted that the I abundance of 3 ppb [57] and the corresponding closure age of 21 Ma after CAI are likely to be a lower limit, so that a ≈ 30-60 Ma possible range, corresponding to a terrestrial I content of 6-13 ppb, is preferred.

*(b) I-Pu-Xe closure age*

The I-Xe closure age given above is in fact a mass balance between the residual amount of radiogenic $^{129}$Xe in the atmosphere and the amount of initial $^{129}$I, with the underlying assumption that the former derived from the latter. In fact, the whole amount of $^{129}$Xe(I) could have been added to a "dead" Earth (that is, an Earth closed after the complete decay of $^{129}$I), by an accreting primitive body formed just after CAIs and having therefore a high $^{129}$Xe(I)/I ratio. In such an extreme case, the "closure" age based on I-Xe would not have any chronological meaning for the Earth. Here it becomes interesting to make use of the I-Pu-Xe system in addition to the I-Xe system. The $(^{129}Xe(I)/^{136}Xe(Pu))_{ATM,CORR}$ ratio depends on two decay constants having contrasted values. Thus this ratio is time-dependent and decreases with age. In the example given above, the ($^{129}$Xe(I)/$^{136}$Xe(Pu))$_{ATM,CORR}$ ratio should yield a very young age, contrary to the I-Xe chronometer. In other words, the two chronometers all-together have the faculty to give, or not, "concordant" closure ages.

The atmospheric $^{129}$Xe(I)/$^{136}$Xe(Pu) ratio, corrected for Xe escape, becomes 21.7 instead of 4.6 (Table 2), yielding a closure age of 34 Ma using Eqn. 2, with a possible range of 13-58 Ma. However, the lower time limit is based again on an unrealistically low terrestrial amount of iodine. This range is "concordant" with the I-Xe range of 30-60 Ma and both methods converge to a revised closure age of $\approx 40_{-10}^{+20}$ Ma for the atmosphere, instead of about 100 Ma previously thought (Fig. 3). Interestingly, the $^{129}$Xe(I)/$^{136}$Xe(Pu) ratio has also been estimated for the mantle from the precise Xe isotope analysis of mantle-derived $CO_2$-rich gases [32]. Estimates of this ratio vary between 20 and 64, depending on parameters (e.g. I/Pu/U ratios, primordial Xe isotopic composition) chosen to derive them. This range of values leads to a possible interval of 20-50 Ma for a major episode of mantle degassing [32]. Thus it is tempting to link such an event to a major catastrophic impact that affected the interior of the proto-Earth and its atmosphere and that led to the formation of the Moon. However, it must also be noted that Xe isotope data for mid-ocean ridge basalts seem to define lower $^{129}$Xe(I)/$^{136}$Xe(Pu) ratios around 6-10 [4,5], that could represent another giant event [32], or, in our opinion, fractionation of volatile iodine with respect to refractory plutonium/uranium for the mantle source of MORBs, during Earth's building episodes.

The range of closure ages (30-60 Ma after CAI) is consistent with Hf-W model ages for the Moon's formation predicting an ancient fractionation (25-50 Ma) from the W isotope difference between chondrites and Earth [65–68]. However, it is only marginally consistent with those based on the similarity of the W isotopes ratios between Earth and Moon, that requires the Moon formation event to have taken place $\geq$ 60 Ma after CAIs [69]. The former case would reinforce the link between the last episode of major loss of the atmosphere and the formation of the Moon.

*(c) Physical significance of closure ages and the epoch of the formation of the Earth-Moon system*

The model presented here permits to correct the xenon budget of the Earth's atmosphere for long-term escape to space. This correction leads to a reconstruction of the xenon isotopic composition of the atmosphere just after the last episode of major loss. After correction, the I-Xe and the I-Pu-Xe systems converge towards a common range of closure ages within 30-60 Ma after CAI, suggesting that they have indeed recorded common catastrophic event(s). In fact, the closure ages estimated here probably result of the integration of a suite of events occurring during the accretion and where primitive and differentiated bodies with variable volatile contents contributed to the building of the proto-Earth.

The fraction of the proto-atmosphere removed during a giant impact is debated [45,70]. New modelling results based on the Moon-forming scenario with a fast-spinning Earth [71] suggest that a large part (if not the whole) of the atmosphere could have been removed during this major event [72]. Our reconstructed Xe budget indicates that about 23 % of the total $^{129}$Xe(I) produced by terrestrial $^{129}$I was left in the atmosphere after completion of the Earth. The possibility that ≈ 77 % of initial volatile elements would have been lost before atmosphere's closure is not unreasonable in regard to possible atmospheric erosion during terrestrial accretion [45]. This mass balance implies that the proto-Earth could have hosted a factor of ≈ 4 more volatiles than at present, if these volatiles were supplied before the last giant impact event. Since the present-day terrestrial inventory of volatile element is estimated to be equivalent to about 2±1 % of carbonaceous chondrite (CC) material [73], this mass balance suggests that the proto-Earth could have been contributed by up to ≈ 10 % CC material before the last giant impact. Alternatively, terrestrial volatiles could have been supplied from bodies having Xe/I ratios higher than those observed presently in CC (e.g., cometary material?).

The conclusions drawn here have to be taken carefully as there all still too many shady areas in the early accretional history of the Earth. Our model does not pretend to describe such accretional processes, which will require a continuous accretion/degassing simulation. The main result of this study is nevertheless that the age of the terrestrial atmosphere appears closer to ≈ 40 Ma rather than to 100 Ma as previously thought.

## 5. Conclusion

The I-Pu-Xe system gives useful, yet not fully understood, time constraints "on establishment of global chemical inventories" [17] for the Earth and, presumably, for the Moon. After geological loss of Xe from the atmosphere, the closure ages derived from the I-Xe and I-Pu-Xe systematics are more ancient than previously proposed and suggest major forming events around 40 Ma (range 30-60 Ma) after CAI. A more comprehensive approach of this chronology will need to integrate the I-Pu-Xe system into n-body simulations of terrestrial accretion, parameterized with gain and loss of volatiles, especially during the giant impact epoch.

**Acknowledgments.**

David Stevenson and Alex Halliday, organisers of the "Origin of the Moon" meeting, are gratefully acknowledged as well as other participants of the meeting for fruitful discussions. Remarks from Maïa Kuga and Eric Hébrard helped to build this model. We are grateful to Allessandro Morbidelli, Seth Jacobson, and Patrick Michel for sharing exciting ideas on the accretion of terrestrial planets, and to Jamie Gilmour and Sujoy Mukhopadhyay for constructive reviews. This work was supported by the European Research Council under the European Community's Seventh Framework Program (FP7/2007-2013 grant agreement no. 267255 to B.M.). This is CRPG contribution #2308

**Tables**

Table 1: Input parameters of the model

| parameter | | literature | note | adjusted value* |
|---|---|---|---|---|
| $I_{INI}$ | ppb | 3-13 | a | **6.4** |
| $(^{129}I/^{127}I)_{INI}$ | mol/mol | $1.1 \times 10^{-4}$ | b | $1.1 \times 10^{-4}$ |
| $^{238}U_{INI}$ | ppb | 33-41 | c | **40** |
| $(^{244}Pu/^{238}U)_{INI}$ | mol/mol | $6.8 \times 10^{-3}$ | d | $6.8 \times 10^{-3}$ |
| $^{244}Pu_{INI}$ | mol | | e | **$3.2 \times 10^{15}$** |
| $^{130}Xe_{INI}$ | mol | n.d. | f | **$1.44 \times 10^{13}$** |
| $(^{124}Xe/^{130}Xe)_{INI}$ | mol/mol | 0.02928 | g | 0.02928 |
| $(^{126}Xe/^{130}Xe)_{INI}$ | - | 0.02534 | - | 0.02534 |
| $(^{128}Xe/^{130}Xe)_{INI}$ | - | 0.5083 | - | 0.5083 |
| $(^{129}Xe/^{130}Xe)_{INI}$ | - | 6.286 | - | 6.286 |
| $(^{131}Xe/^{130}Xe)_{INI}$ | - | 4.996 | - | 4.996 |
| $(^{132}Xe/^{130}Xe)_{INI}$ | - | 6.047 | - | 6.047 |
| $(^{134}Xe/^{130}Xe)_{INI}$ | - | 2.126 | - | 2.126 |
| $(^{136}Xe/^{130}Xe)_{INI}$ | - | 1.657 | - | 1.657 |
| degassing parameters | | | | |
| $\varphi_1$ | $Ma^{-1}$ | n.d | h | **0.065** |
| $\varphi_2$ | - | n.d | - | **0.0011** |
| $\varphi_3$ | - | n.d | - | **0.000073** |
| escape parameters | | | | |
| b | $Ma^{-1}$ | 2.53 | i | 2.53 |
| c | - | 0.85 | - | 0.85 |
| d | - | n.d | - | **1.2270** |
| instantaneous fractionation | %.amu$^{-1}$ | 1.0-1.5 | j | **1.3** |

notes:

* bold values are adjusted parameters using multiple runs of the model.

a, b: See section 3.(a)

c, d, e: See section 3.(b)

g: See section 3.(c)

h, i, j: See section 3.(d)

Table 2: Reference solution of the model compared to values from the literature

| parameter | dimension | literature | note | reference solution |
|---|---|---|---|---|
| $Xe_{ATM}$ | mol | $1.537 \times 10^{13}$ | a | $1.5184 \times 10^{13}$ |
| $(^{124}Xe/^{130}Xe)_{ATM}$ | mol/mol | 0.02337 | - | 0.0232 |
| $(^{126}Xe/^{130}Xe)_{ATM}$ | - | 0.0218 | - | 0.0217 |
| $(^{128}Xe/^{130}Xe)_{ATM}$ | - | 0.4715 | - | 0.470 |
| $(^{129}Xe/^{130}Xe)_{ATM}$ | - | 6.496 | - | 6.522 |
| $(^{131}Xe/^{130}Xe)_{ATM}$ | - | 5.213 | - | 5.21 |
| $(^{132}Xe/^{130}Xe)_{ATM}$ | - | 6.607 | - | 6.60 |
| $(^{134}Xe/^{130}Xe)_{ATM}$ | - | 2.563 | - | 2.55 |
| $(^{136}Xe/^{130}Xe)_{ATM}$ | - | 2.176 | - | 2.17 |
| $^{129}Xe(I)_{ATM,CORR}$ | mol | | b | $3.71 \times 10^{12}$ |
| $(^{129}Xe(I)/^{136}Xe(Pu))_{ATM}$ | mol/mol | 6.5-7.1 | c | 6.99 |
| $(^{129}Xe(I)/^{136}Xe(Pu))_{ATM,CORR}$ | - | - | d | 21.75 |
| $Xe_{MANT}$ | mol | | | $2.0356 \times 10^{12}$ |
| $(^{124}Xe/^{130}Xe)_{MANT}$ | mol/mol | | | 0.0293 |
| $(^{126}Xe/^{130}Xe)_{MANT}$ | - | | | 0.0253 |
| $(^{128}Xe/^{130}Xe)_{MANT}$ | - | | | 0.508 |
| $(^{129}Xe/^{130}Xe)_{MANT}$ | - | | | 9.00 |
| $(^{131}Xe/^{130}Xe)_{MANT}$ | - | | | 5.18 |
| $(^{132}Xe/^{130}Xe)_{MANT}$ | - | | | 6.73 |
| $(^{134}Xe/^{130}Xe)_{MANT}$ | - | | | 2.86 |
| $(^{136}Xe/^{130}Xe)_{MANT}$ | - | | | 2.50 |

notes:

*Bold values are ajusted with multiple runs of the model (see Section 3.4).

a- The total inventory and isotopic ratios of Xe are from [31].

b- Amount of $^{129}$Xe in the atmosphere coming from the decay of $^{129}$I and corrected for loss.

c- Ratio of radioactivity products uncorrected for loss from [74].

d- Ratio corrected for loss.

**Figure captions**

Fig. 1: Relationship between the evolution of the solar EUV flux with time and the progressive isotopic fractionation of atmospheric xenon. (a) Evolution of the Solar flux with time. Figure and data are from [41]. The wavelength of ionization of Xe atoms corresponds to the range 920-1200 Å. (b) Progressive isotopic fractionation of atmospheric Xe from cosmochemical components [10] to modern atmosphere [31]. Ancient rocks record intermediate isotopic compositions: 1 Ga-old barites [35,36], Quartz samples [37,38], Proterozoic deep fracture fluids [39].

Fig. 2: (a) Schematic explanation of the model and evolution of the budget of atmospheric Xe over time. The model is built with three boxes: solid Earth, atmosphere and space. Some of Xe isotopes are produced by radioactive decay of $^{129}$I, $^{244}$Pu and $^{238}$U (see text). After a time interval Δt (40 Ma, one outcome of the model) the Earth begins to retain its volatile elements and accumulates xenon degassed from the Solid Earth to the atmosphere without isotopic fractionation. Xenon atoms escape from the atmosphere to the outer space with isotopic fractionation. The evolution of the budget of atmospheric xenon shows the progressive escape of Xe atoms with time. The escape lasts until the end of the Archean eon (t=2 Ga). At this time, the abundance has almost reached the current abundance of xenon in the atmosphere. Further degassing permits to complete the amount of xenon atoms. (b) Isotopic spectrum of xenon relative to the current isotopic composition of the Earth's atmosphere using $^{130}$Xe as a reference isotope. Xe-U is the starting isotopic composition (circles). The fractionated Archean atmosphere (around 1%.amu$^{-1}$) is shown with squares and the "artificial" current isotopic composition of the reference solution is shown with stars. The current isotopic composition is reproduced within 0.7% or better.

Fig. 3: (a) Evolution of the atmospheric content of $^{129}$Xe derived from the decay of $^{129}$I with time. The non-corrected amount gave a closure age of 98 Ma for the I-Xe system. After correction for subsequent loss, the age becomes 41 Ma. (b) Evolution with time of the ratio of radioactive products in the atmosphere ($^{129}$Xe(I) and $^{136}$Xe(Pu)). The non-corrected ratio gave a closure age of 66 Ma for the I-Pu-Xe system. After correction, the age becomes 34 Ma in agreement with the time of closure of the I-Xe system and with the closure age of the mantle given by the mantle samples [4,32].

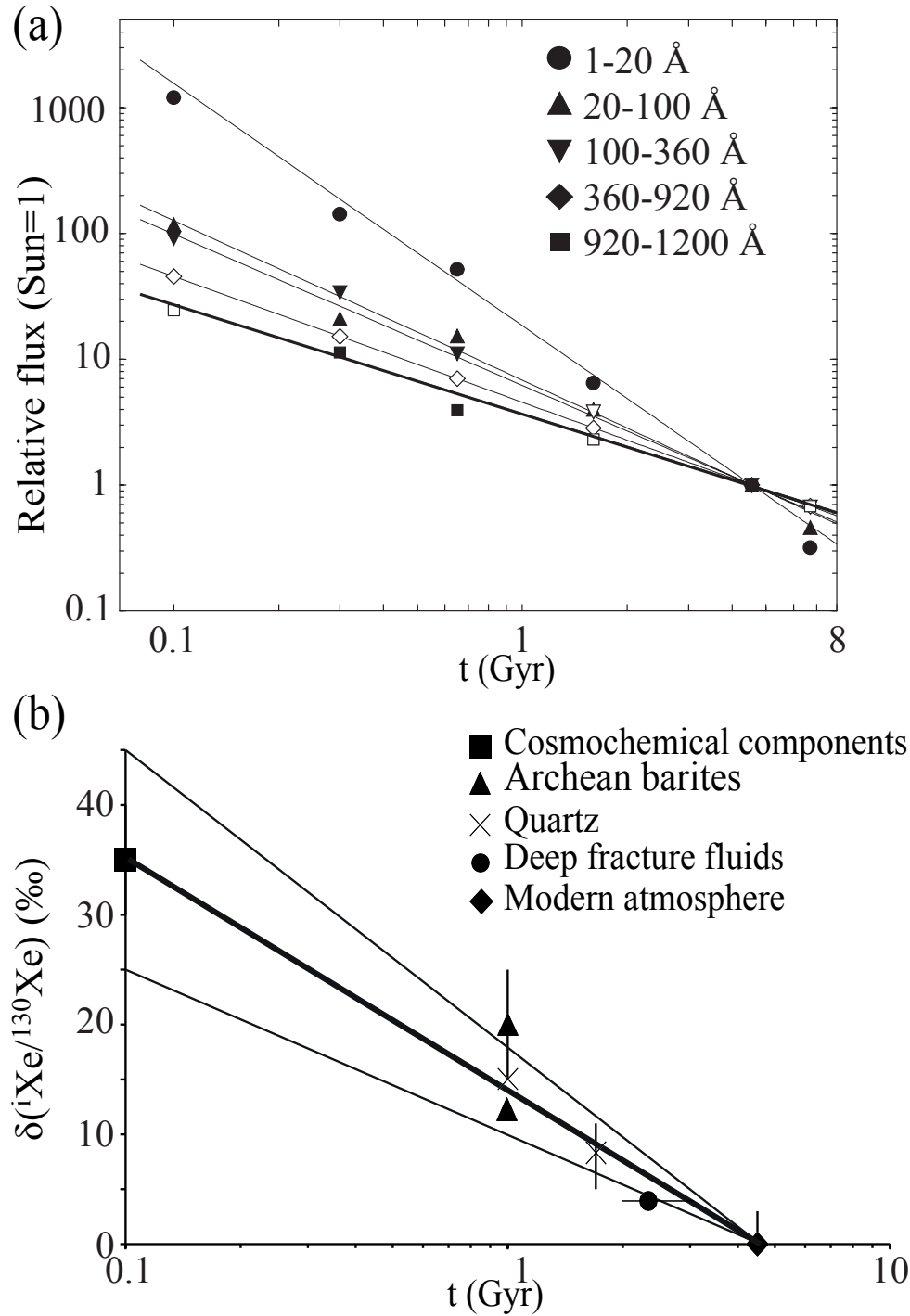

**Figure 1**

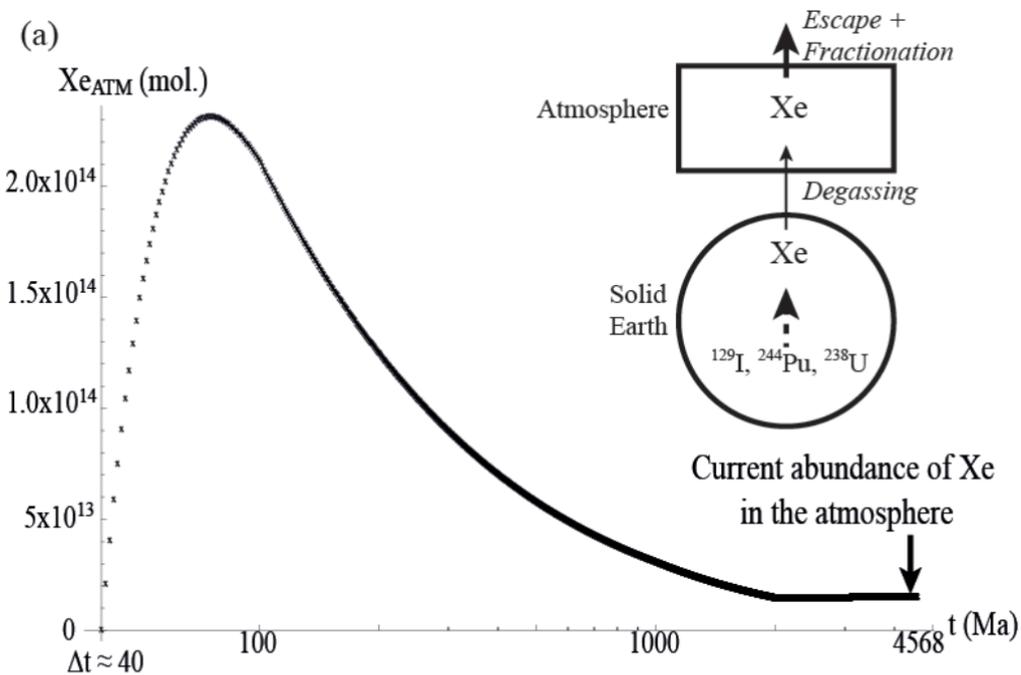

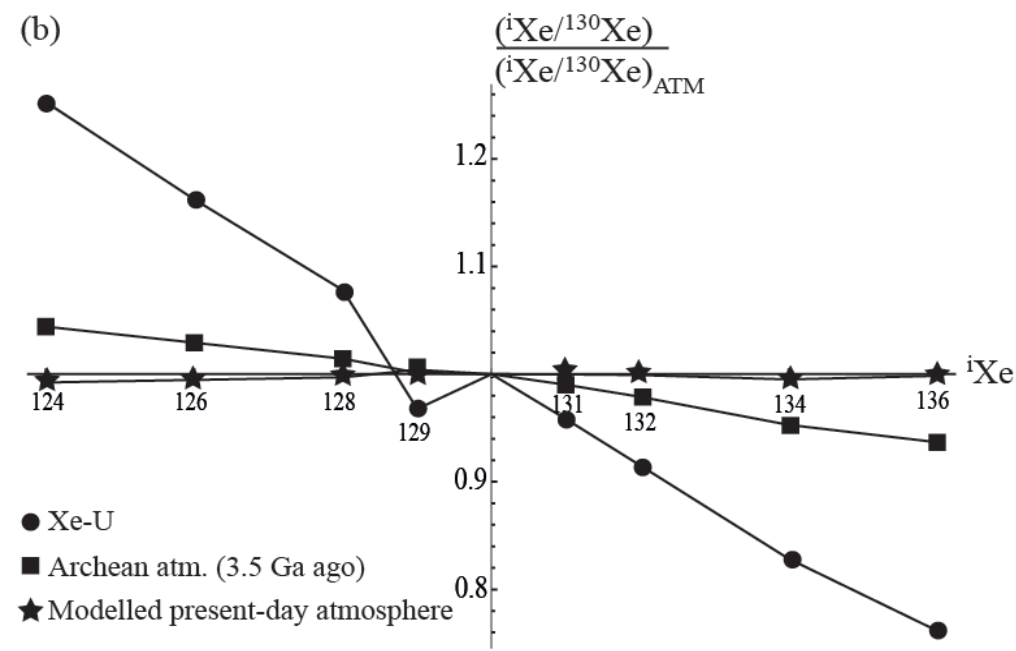

**Figure 2**

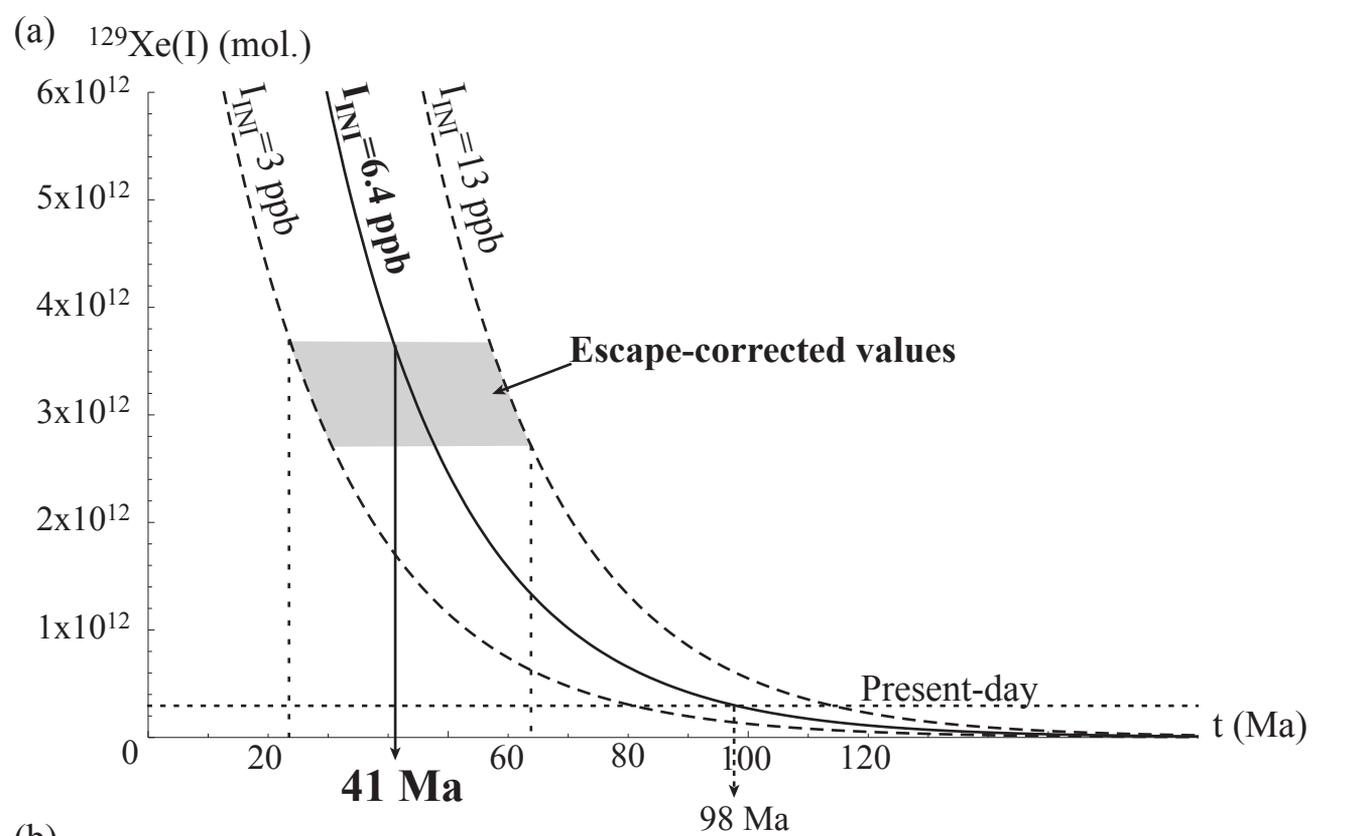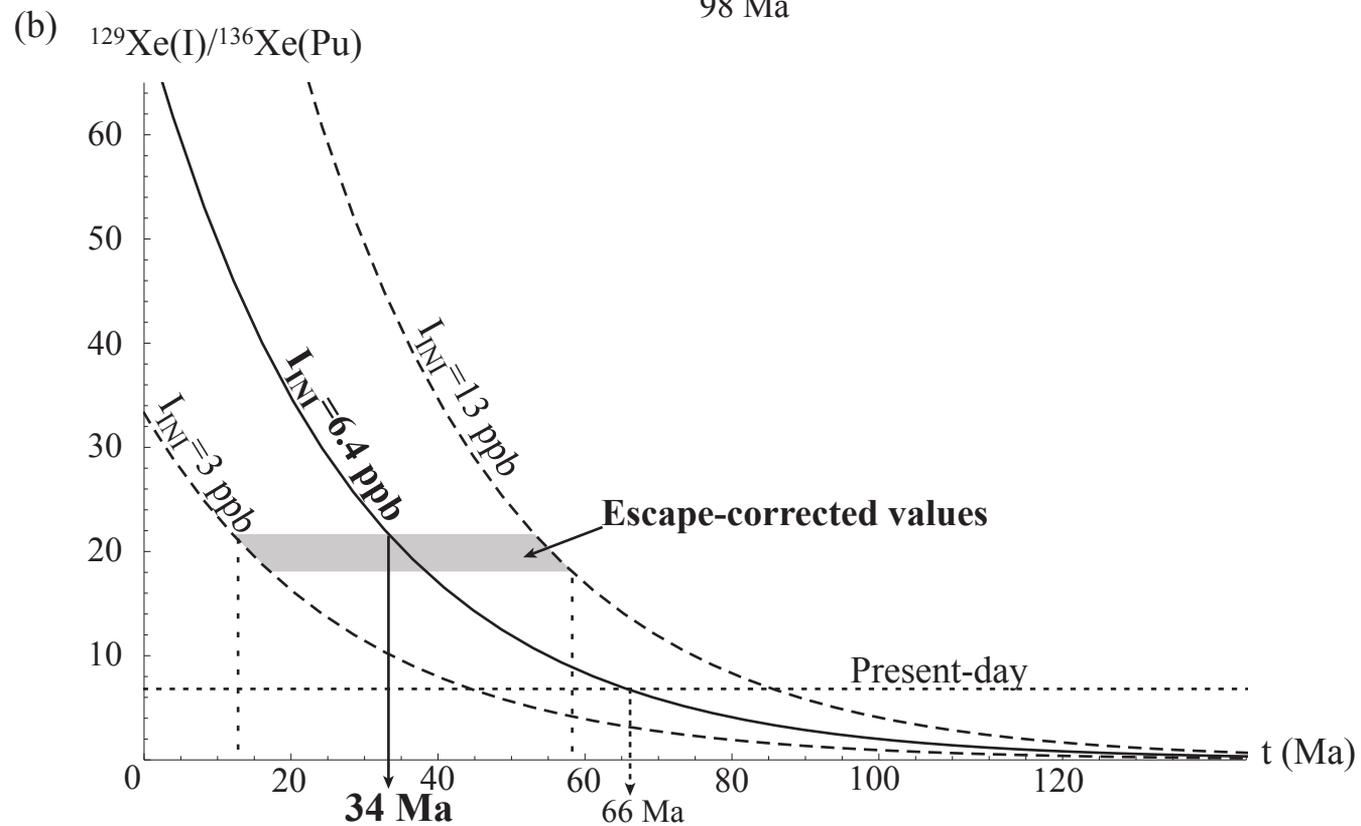

**Figure 3**